\documentclass[pre,showpacs,showkeys,twocolumn,10pt]{revtex4-1}
\usepackage{amsfonts,bbold,amsmath,amssymb,graphicx,epstopdf,verbatim,dsfont,color}
\usepackage[english]{babel}

\newcommand{\tr}{{\rm Tr}}

\begin{document}

\title{Stationary ensemble approximations of dynamic quantum states: 
\\ Optimizing the Generalized Gibbs Ensemble}
\author{Dries Sels}
\affiliation{TQC, Universiteit Antwerpen, Universiteitsplein 1, B-2610 Antwerpen, Belgium}
\author{Michiel Wouters}
\affiliation{TQC, Universiteit Antwerpen, Universiteitsplein 1, B-2610 Antwerpen, Belgium}
\date{\today}

\begin{abstract}
We reconsider the non-equilibrium dynamics of closed quantum systems. In particular we focus on the thermalization of integrable systems. Here we show how the generalized Gibbs Ensemble (GGE) can be constructed as the best approximation to the time dependent density matrix. Our procedure allows for a systematic construction of the GGE by a constrained minimization of the distance between the latter and the true state. Moreover, we show that the entropy of the GGE is a direct measure for the quality of the approximation. We apply our method to a quenched hard core bose gas. In contrast to the standard GGE, our correlated GGE properly describes the higher order correlation functions.
\end{abstract}
\maketitle

\section{Introduction}
Thermalization is a fundamental feature of closed many-body systems on which much progress has been made in recent years~\cite{Rigol2, ReimannRealistic, ReimannKastner, Tasaki, GoldsteinTypicality, Popescu, Linden, ReimannOpen, ShortFinite}. After the early pioneering work by von Neumann \cite{neum,gold_neum}, the subject was put aside for a long time and thermalization was mainly understood as the result of interactions with an environment. The advent of closed quantum systems with a high degree of controllability, in particular ultracold atomic gases \cite{bloch,Schmiedmayer}, has revived the interest in this subject. The implementation of integrable \cite{weiss} cold atom systems has also sparked the interest in understanding the role of integrability in thermalization.

Here we are concerned with the statistical description of the system after a long time. For generic systems, one expects the observables to thermalize such that their stationary value can be obtained from the the standard Boltzmann-Gibbs ensemble: $\rho_{\rm eq}=\exp(-\beta \hat H)$. For integrable systems, in contrast, the canonical Boltzmann-Gibbs ensemble may even fail to describe elementary properties such as the momentum distribution. The strong modification of the stationary state is due to the existence of an additional set of conserved quantities $\hat I_\ell$. A particularly striking example is given by one dimensional system of hard core bosons which has conserved occupation numbers~\cite{weiss, Rigol1, Rousseau}.  

The formalism to generalize the Boltzmann-Gibbs ensemble was proposed by Jaynes~\cite{Jaynes}, who pointed out that statistical physics can be seen as statistical inference and an ensemble as the least biased estimate possible on the given information. By including the conserved observables in our information, the Boltzmann-Gibbs ensemble is generalized to $\rho_{\rm eq}=\exp\left(- \sum_\ell \alpha_\ell \hat I_\ell\right)$, where the Lagrange multipliers  $\alpha_\ell$ have to be adjusted for the ensemble to predict the correct expectation values. 
The momentum distribution of quenched hard core bosons has been successfully described by such a generalized Gibbs Ensemble (GGE), that takes into account the conserved occupation numbers \cite{Rigol1}. This clearly shows the relevance of the conserved quantities in constructing ensembles for quantum systems. 

Unfortunately, it is not always clear which conserved quantities have to be taken into account~\cite{WoutersCaux,Takacs,Goldstein,Essler,Fagotti1}. Actually, the linearity of quantum mechanics implies that the set of conserved quantities contains all projectors $\hat P_n$ on eigenstates of the Hamiltonian $H=\sum_n \epsilon_n \hat P_n$. The number of these projection operators scales with the size of Hilbert space and is exponential in the physical system size, e.g. particle number. The ensemble that takes into account all expectation values of these projectors is the so-called diagonal ensemble \cite{polkov_diag}. In order to construct this ensemble, one needs the exponentially many diagonal elements of the density matrix in the eigenbasis of the Hamiltonian. 

Clearly, the diagonal ensemble contains much more information than the GGE. The fact that more information than the one included in the GGE may be relevant is easy to see: if $\hat I_\ell$ is conserved then also $\hat I_\ell \hat I_{\ell'}$ is conserved. This means that initially non-zero correlations between the different integrals of motion are conserved, where the GGE predicts no correlations between them. This fact has led to criticism on the validity of the GGE \cite{Gangardt}. Furthermore, an extensive difference in the entropy between the diagonal and generalized Gibbs ensembles was reported by several authors \cite{Santos,Caux,Fagotti2}, which implies the two ensembles are macroscopically distinguishable.

An elementary question thus arises: which information is essential to include in the generalized Gibbs ensemble in order to get the correct entropy and correct expectation values for all the observables? The `statistical inference approach' by Jaynes is not very useful in this respect. It would simply tell us that we should use the diagonal elements of the density matrix if we have access to them. This does not answer the question whether it is possible to construct a more economical representation of the density matrix that does not scale exponentially with system size.
\section{Coherence and the minimal distance}
In order to address this problem, we take a slightly different view on the task of statistical mechanics. Instead of maximizing the entropy subject to constraints, we wish to minimize the average distance between a stationary trial ensemble $\hat{\sigma}$ and the true state $\hat{\rho}(t)$. This formulation as an optimisation problem will allow us to systematically improve the ensemble by including more observables when needed.
We thus set out to minimize the average distance between a stationary trial ensemble $\hat{\sigma}$ and the state $\hat{\rho}(t)$ over time
\begin{equation}
\overline{D}=\lim_{T\rightarrow \infty}\frac{1}{T} \int_0^T {\rm d}t D\left(\hat{\rho}(t)\Vert \hat{\sigma} \right). 
\label{eq:distanceFull}
\end{equation}

As a measure for the distance between density matrices  we adopt the Kullback-Leibler (KL) distance~\cite{KL} (also known as relative entropy). The reason for this is twofold. First of all, when supplied with canonical distributions, the KL distance can immediately be related to thermodynamic quantities such as entropy and work. For an example of the latter we refer to~\cite{vdB}. Secondly, it will turn out that the minimal KL distance is directly related to the amount of coherence in the true state.

After time averaging the KL distance becomes 
\begin{equation}
\overline{D}=-S\left( \hat{\rho}_0\right)-\tr\left[\hat{\rho}_d \log\left(\hat{\sigma} \right)  \right], 
\label{eq:avdis}
\end{equation}
where $S(\hat \rho) = - \tr (\hat \rho \ln \hat \rho)$ denotes the von Neumann entropy, $\hat{\rho}_{\rm d}$ the diagonal ensemble (DE) \cite{polkov_diag}, i.e. 
\begin{equation}
\hat{\rho}_{\rm d}=\lim_{T\rightarrow \infty}\frac{1}{T} \int_0^T {\rm d}t \; \hat{\rho}(t),
\nonumber
\end{equation}
and $\hat{\rho}_0=\hat{\rho}(0)$. Note that we have simplified expression \eqref{eq:avdis} by making use of the fact that the von Neumann entropy is conserved under unitary time evolution. Next, one readily finds that the distance is minimized for $\hat{\sigma}=\hat{\rho}_{\rm d}$ with the minimal distance equal to 
\begin{equation}
\overline{D}_{\rm min}=S\left( \hat{\rho}_{\rm d}\right)-S\left( \hat{\rho}_0\right)
\end{equation}
The diagonal ensemble thus always represents the optimal stationary approximation of the true state and the diagonal entropy quantifies its distance to the instantaneous density matrix, the latter moreover quantifies the amount of coherence of the true state~\cite{Plenio}.
In the case of a pure state, for which $S\left( \hat{\rho}_0\right)=0$, the diagonal entropy is equal to the Kullback-Leibler distance between the instantaneous state $\hat \rho(t)$ and the time independent approximation $\hat \sigma$. Entropy is then identified as the information that is lost by describing a time dependent system by a time independent density matrix. We thus have a straightforward information theoretic interpretation of the entropy of a pure state without needing to separate the system in subsystems, as required for the definition of the entanglement entropy \cite{Popescu, Linden}.
 
By unconstrained minimization, we have recovered the diagonal entropy, that requires the specification of a number of parameters that is exponential in system size. Let us now address the problem of constructing a stationary matrix that captures the relevant physics while being more economical in the number of parameters.
Consider therefore the ansatz
\begin{equation}
\hat{\sigma}=Z_{\sigma}^{-1}\exp\left[-\sum_\ell \lambda_\ell \hat{A}_\ell \right], 
\label{eq:ansatz}
\end{equation}
where $\left\lbrace \hat{A}_\ell \right\rbrace$  is a set of operators, which we leave unspecified for the moment. Minimization of the distance~\eqref{eq:distanceFull} with respect to the set $\left\lbrace \lambda_\ell \right\rbrace$ yields the condition that
\begin{equation}
\tr \left[ \hat{A}_\ell\hat{\sigma} \right]= \tr \left[ \hat{A}_\ell \hat{\rho}_{\rm d} \right],
\label{eq:equalExpec}
\end{equation}
and the minimal distance is consequently equal to 
\begin{equation}
\overline{D}_{\rm min}=S\left( \hat{\sigma}\right)-S\left( \hat{\rho}_0\right).
\end{equation}
Again the difference in entropy between the trial and the initial state quantifies the distance between the two. Furthermore, because the Kullback-Leibler distance is always positive, the entropy of the canonical trial is always bigger than that in the diagonal ensemble and the entropy difference is equal to the distance between the two, i.e. $ D\left(\hat{\rho}_{\rm d}\Vert \hat{\sigma} \right)=S\left( \hat{\sigma}\right)-S\left( \hat{\rho}_{\rm d}\right)$. The total distance can thus be decomposed in two components, the amount of coherence of the state and the inability of the ensemble $\hat{\sigma}$ to describe the diagonal ensemble. 

It is important to note that, even though our ansatz \eqref{eq:ansatz} and the conditions \eqref{eq:equalExpec} are the same as in Jaynes work, our interpretation is very different.
In our approach, the entropy has to be {\em minimized} under the constraints \eqref{eq:equalExpec} by using the best set of observables  $\left\lbrace \hat{A}_\ell \right\rbrace$. In Jaynes's approach on the other hand, the entropy is maximized for a set of observables that is fixed beforehand. This difference is important for the following reason. A lower value of the entropy can be reached by including more observables. It therefore makes sense to construct the optimal density matrix \eqref{eq:ansatz} with as few observables as possible. This procedure then allows to identify the physically relevant observables. Jaynes' maximization of the entropy does not offer this principle, because the maximum entropy is reached when no observables are used (equal occupation of all energy eigenstates).

The physical importance of the Kullback-Leibler distance is brought forward by Pinsker's inequality \cite{Pinsker}, combined with the operational interpretation of the trace distance \cite{Wilde}.  These guarantee that the difference in any projection operator valued measurement (POVM) is bound from above by the Kullback-Leibler distance:
\begin{equation}
D\left(\hat{\rho}_{\rm d}\Vert \hat{\sigma} \right) \geq  \frac{1}{2} \max_{B} \left\lbrace  \left(  \sum_\ell \left\rvert \tr \left[ \hat{B}_\ell\hat{\sigma} \right]- \tr \left[ \hat{B}_\ell \hat{\rho}_{\rm d} \right] \right\rvert \right)^2 \right\rbrace , 
\label{eq:TraceM}
\end{equation}
where the maximum is taken over all sets of operators $\left\lbrace B \right\rbrace  $  that form a resolution of the identity.
Consequently, if a set of operators $\{ \hat A_\ell \}$ exists that makes the distance between the GGE and the diagonal ensembles small, {\em all} POVMs show small differences between the two ensembles.

It is important to note that because of the optimum condition \eqref{eq:equalExpec}, the operators that are included in the definition of the GGE do not show any deviation between the diagonal and GGE ensembles. 
This suggests an iterative construction of the GGE, where operators that show the largest deviation between the diagonal and GGE ensembles are included in the set $\{ \hat A_\ell \}$. Depending on the system, the convergence of this procedure will yield a different number and different kinds of operators to be included in the GGE. 

This discussion is valid for both integrable and non-integrable quantum systems. For the latter there is compelling numerical evidence \cite{Santos, RigolThermCluster} that they thermalize, such that the standard statistical mechanics ensembles  provide a good approximation to the diagonal ensemble for what concerns the description of physical (few-body) observables. It is then sufficient to include the Hamiltonian, number operator and possibly a few other operators related to global symmetries in the trial ensemble. Integrable quantum systems behave quite differently. Any further discussion on how to concretely implement the above procedure must of course be based on a specific model. 

\section{Integrable systems: an example}
As discussed above, additional conserved quantities are essential to be included in the construction of the ensemble. In the following, we will consider non interacting fermions because of their experimental relevance \cite{bloch} and because many other 1D models can be mapped via a Jordan-Wigner transformation on free fermions, e.g. transverse field Ising model and hardcore bosons \cite{Rousseau}.

More specifically we numerically study a quench in one dimensional non-interacting lattice fermions by switching on an additional periodic potential. The system under consideration is identical to that in \cite{Santos}, where it was used to describe hard-core bosons. The Hamiltonian reads
\begin{equation}
H=-\sum_{j=1}^N t \left( \hat{c}^{\dagger}_{j+1}\hat{c}_j + {\rm h.c.} \right) +J \sum_{j=1}^N \cos \left(\frac{2 \pi}{\lambda} j \right) \hat{c}^{\dagger}_{j}\hat{c}_j.
\label{eq:Hquench}
\end{equation}
The period $\lambda=5$, and tunneling rate $t=1$ are used throughout the article and we consider the strength of the potential $J$ as the quench parameter. In all quenches we start in the ground state~\cite{numbeta} of~\eqref{eq:Hquench} with $J=0$ and quench to different values of $J$.

Surprisingly, it was found in Ref. \cite{Santos} that in this simple system, the entropies calculated in the GGE and diagonal ensembles show an extensive difference. According to~\eqref{eq:TraceM}, this means that there may be macroscopic observables that show large differences between the diagonal ensemble and the GGE. Indeed, it was pointed out by Gangardt and Pulstilnik \cite{Gangardt} that correlations in momentum space are wrongly described by the GGE, casting doubt on its validity. 

\begin{figure}[t]
\includegraphics[width=\columnwidth]{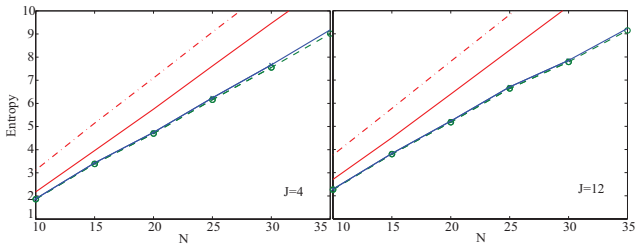}
\caption{(Color online)~The entropy as a function of system size for quenches from $J=0$ to $J=4$ (left hand panel) and $J=12$ (right hand panel). The red dash-dotted line shows the entropy in the GGE; red full line is the GCE~\cite{noteGCE}. The green dashed line and circles shows the diagonal entropy and the blue line with crosses is the CGGE entropy.}
\label{fig:Entropy}
\end{figure}

The reason for the failure of the GGE can be seen by noting that all information about the time-dependent state $\hat{\rho}(t)$ is contained in the one-body density matrix 
\begin{equation}
g^{(1)}_{ij}(t)=\hat{a}^{\dagger}_{i}(t)\hat{a}_j(t),
\label{eq:rho_1}
\end{equation}
where $\hat{a}_j$ annihilates a particle in a post-quench eigenstate, i.e. Hamiltonian (\ref{eq:Hquench}) can be written as $H=\sum_j \epsilon_j \hat{a}^{\dagger}_{j}\hat{a}_j$. Only the diagonal, $g^{(1)}_{ii}$, is conserved over time such that it is the only part of the operator with a non-trivial time averaged expectation value. It should therefore be included in the trial ensemble. If one only includes these operators one immediately arrives at the standard generalized Gibbs ensemble
\begin{equation}
\hat{\rho}_{ \rm GGE}=Z^{-1} \exp\left[ -\sum_j \lambda_j \hat{a}^{\dagger}_{j}\hat{a}_j  \right] .
\end{equation}
All off-diagonal elements vanish because their phases rotate at a frequency $\epsilon_i-\epsilon_j$. However, the amplitude of the elements is conserved over time and that information is completely lost in the GGE while retained in the diagonal ensemble. As a consequence the GGE predicts a wrong value for the 2-body correlation:
\begin{equation}
\tr \left[ \hat{a}^{\dagger}_{j}\hat{a}^{\dagger}_{i}\hat{a}_i\hat{a}_j \left(\hat{\rho}_{\rm GGE}-\hat{\rho}_{\rm d} \right) \right]= \left\rvert g^{(1)}_{ij} \right\rvert^2-n_i^2 \delta_{i,j}. 
\end{equation}
As the difference is significant we must include a 2-body interaction term in the trial ensemble to remove the discrepancy between the GGE and the DE. We therefore propose the improved correlated generalized Gibbs ensemble (CGGE)
\begin{equation}
\hat{\rho}_{ \rm CGGE}=Z^{-1} \exp\left[ -\sum_j \lambda_j \hat{a}^{\dagger}_{j}\hat{a}_j- \sum_{i,j} V_{i,j} \hat{a}^{\dagger}_{j}\hat{a}^{\dagger}_{i}\hat{a}_i\hat{a}_j  \right],
\label{eq:cgge}
\end{equation}
which includes a density-density interaction $V_{i,j}$. In order to compare all ensembles we calculate their entropy which is depicted in Fig.~\ref{fig:Entropy}.
It is immediately clear that in contrast to the GGE, the CGGE has the same slope as the diagonal ensemble. This implies that, while there is an extensive difference in entropy between the GGE and the diagonal ensemble, the difference between CGGE and diagonal ensemble is sub-extensive. Moreover, both entropies are so close that they put a strong bound on the trace distance between the two ensembles.  
\begin{figure}[h]
\includegraphics[width=\columnwidth]{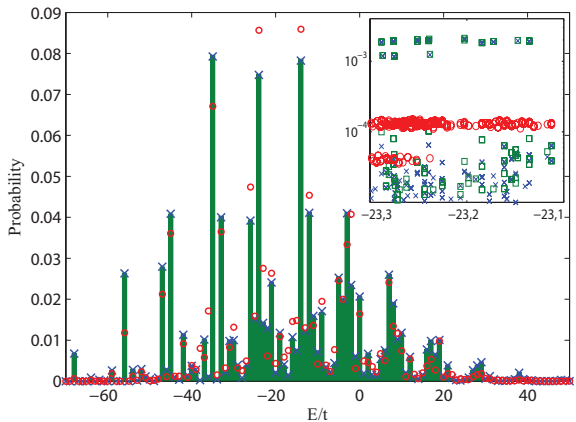}
\caption{(Color online)~The coarse grained ($\Delta E /t=1$) energy distribution for the $J=12$, $N=30$ quench. The green bars show the exact distribution. Red circles are the GCE prediction and the blue crosses is the present result. The inset shows some non-course grained matrix elements on a semi-logarithmic scale; blue crosses are the CGGE, red circles the GCE and green boxes the DE.}
\label{fig:EnergyDis}
\end{figure} 
This implies that almost all observables are perfectly described in the CGGE. The CGGE should therefore recover the diagonal probabilities with great accuracy. 
At first sight, it could seem surprising that the proper ensemble for quenched free fermions is non-Gaussian, so that correlation functions cannot be computed with Wick's theorem. This is however a simple consequence of the mathematical fact that the time average of a time-dependent Gaussian density matrix is not Gaussian. In contrast the GGE is Gaussian which makes it significantly easier to compute than the CGGE. The present situation however shows that the virtue of being easy compute also makes it fail to describe more complicated observables.

In Fig.~\ref{fig:EnergyDis} we compare the different ensembles directly by plotting the diagonal elements for the density matrix. In order to show the whole interval of occupied states, we have made a coarse graining of the energy with $\Delta E =t$. The CGGE coincides well with the diagonal ensemble and both are indistinguishable by eye. In contrast, the prediction of the GGE is rather poor. The total variation distance between the coarse grained GGE and the DE is 0.28, where the distance between the CGGE and DE is only 0.01.
The inset in Fig.~\ref{fig:EnergyDis} shows the probabilities of individual eigenstates over a small energy interval. The CGGE correctly captures all the probabilities of the significantly occupied states, while the GGE does not correctly predict any occupation of the individual eigenstates. In fact for $J=12$ and $N=30$ the trace distance $D_1\left(\hat{\rho}_{\rm d} , \hat{\rho}_{\rm  GCE} \right)=0.867$, which is still much larger than the distance of 0.28 that we obtained after coarse graining in energy.
The trace distance between the DE and the CGGE is only $D_1\left(\hat{\rho}_{\rm d} , \hat{\rho}_{\rm CGGE} \right)=0.028$. Note that the trace distance is defined as
\begin{equation*}
D_1\left(\hat{\rho} , \hat{\sigma} \right)=\frac{1}{2}\tr  \left\vert \hat{\rho}- \hat{\sigma}  \right\vert 
\end{equation*}
It is intimately related to the problem of distinguishing states. In fact the average success probability of distinguishing two, a priori equally likely, states $\hat{\rho}$ and $ \hat{\sigma}$, by an optimal measurement is~\cite{Wilde}~:
\begin{equation*}
P_{\rm succes}=\frac{1}{2}+\frac{1}{2}D_1\left(\hat{\rho} , \hat{\sigma} \right)
\end{equation*}
The small trace distance between the CGGE and DE thus means that they are hardly distinguishable by any POVM. All observables, including higher order correlation functions are consequently correctly described by the CGGE.

\begin{figure}[t]
\includegraphics[width=\columnwidth]{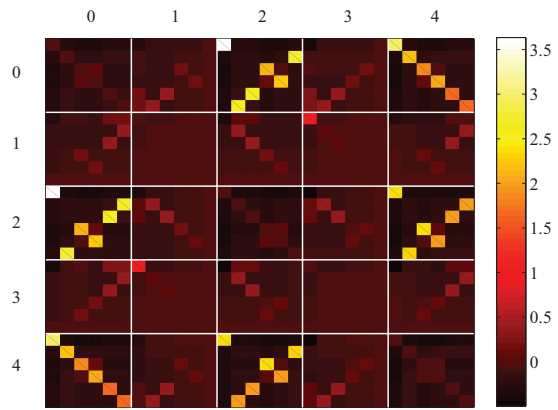}
\caption{(Color online)~The interaction potential $V_{i,j}$ for the $J=12$, $N=30$ quench. Eigenenergies are ascending from top to bottom and left to right. The white lines indicate the different Bloch bands in the reduced Brillouin zone.}
\label{fig:Vij}
\end{figure}

The associated interaction potential $V_{i,j}$ in \eqref{eq:cgge} is depicted in Fig.~\ref{fig:Vij}. As expected, its diagonal is zero, as the diagonal correlations were already properly predicted by the GGE. In fact the overall intra-band interactions are weak. Furthermore, the interaction between bands with odd quantum number, i.e. $\bf{V}_{11}, \bf{V}_{33}, \bf{V}_{13}, \bf{V}_{31} $, are extremely small and tend to zero with increasing $J/t$. This can immediately be understood from the nature of the bands. While the even bands are particle-like, the odd bands are hole-like and can thus not be occupied in the high $J/t$ limit. The interaction is dominated by inter-band repulsion in the electronic bands.

\section{Conclusion}
In conclusion we have formulated the construction of the generalized Gibbs ensemble as an optimization problem. Entropy is interpreted in our approach as the information that is lost by approximating the time dependent state by a stationary density matrix. The diagonal ensemble was shown to be the optimal ensemble when the evolution time tends to infinity. Other ensembles that depend on less parameters can be constructed by minimizing the Kullback-Leibler distance to the diagonal ensemble. 
We have applied these ideas to a quench in a non interacting Fermi system, which shows significant discrepancies between the diagonal and standard GGE. We have constructed a correlated GGE that shows excellent agreement with the diagonal ensemble, providing accurate predictions for all observables and not only for the single particle distribution. 

Since our approach does not rely on any assumption of thermal equilibrium, it could also be applied to construct GGEs for systems out of equilibrium, such as periodically driven systems  \cite{dalessio} or driven-dissipative many body systems \cite{iacrmp}.

\begin{acknowledgments}
We thank J. Tempere for useful comments on the manuscript. D.S. acknowledges support of the FWO as post-doctoral fellow of the Research Foundation - Flanders. M. W. acknowledges financial support from the FWO Odysseus program.
\end{acknowledgments}


\begin{thebibliography}{99}
\bibitem{Pinsker} Pinsker's inequality implies that the trace distance $ D_1\left(\hat{\rho}_{\rm d} , \hat{\sigma} \right)\leq \sqrt{\frac{1}{2} D\left(\hat{\rho}_{\rm d}\Vert \hat{\sigma} \right)} $
\bibitem{noteGCE} The GCE is the canonical version of the GGE, i.e. $S_{GCE}=\tr_N\left[\hat{\rho}_{GCE} \log\left(\hat{\rho}_{GCE} \right)  \right]$ where the trace is only over states with the same particle number.
\bibitem{numbeta} Numerically we start from a canonical state with $\beta t=55$.
\end{thebibliography}
\end{document}